\def\year{2018}\relax
\documentclass[letterpaper]{article} 
\usepackage{aaai18}  
\usepackage{times}  
\usepackage{helvet}  
\usepackage{courier}  
\usepackage{url}  
\usepackage{graphicx}  
\usepackage{algorithm}
\usepackage{algorithmic}
\usepackage{subfigure}
\usepackage{color}
\usepackage{refcount}
\usepackage{paralist}
\usepackage{mathtools}
\usepackage{array} 
\usepackage{mdwlist} 
\usepackage{breqn}
\usepackage{multirow}
\usepackage[font=small,labelfont=bf]{caption}
\usepackage{etoolbox}
\RequirePackage{tcolorbox}
\definecolor{brightpink}{rgb}{1.0, 0.0, 0.5}
\definecolor{darkwhite}{rgb}{.85, 0.85, 0.85}
\definecolor{purple}{rgb}{0.58, 0.44, 0.86}
\definecolor{metal}{rgb}{0.43, 0.5, 0.5}
\definecolor{silver}{rgb}{0.75, 0.75, 0.75}
\definecolor{wooden}{rgb}{0.87, 0.72, 0.53}
\usepackage{colortbl}

\begin{document}
%
\title{``Woman-Metal-White vs Man-Dress-Shorts'':\\ Combining Social, Temporal and Image Signals to\\ Understand Popularity of Pinterest Fashion Boards}
\author{Suman Kalyan Maity\thanks{Most of the work was done when the author was at IIT Kharagpur, India.}$^\S$, Anshit Chaudhary$^{*\dag}$ and Animesh Mukherjee$^\ddag$\\
$^\S$Northwestern University; $^\ddag$Dept. of CSE, IIT Kharagpur, India\\
}

\maketitle
\begin{abstract}
Pinterest is a popular photo sharing website. Fashion is one the most popular and content generating category on this platform. Most of the popular fashion brands and designers use boards on Pinterest for showcasing their products. However, the characteristics of popular fashion boards are not well-known. These characteristics can be used for predicting popularity of a nascent board. Further, newly formed boards can organize their content in a way similar to the popular fashion boards to garner enhanced popularity. What properties on these fashion boards determine their popularity? Can these properties be systematically quantified? In this paper, we show how \textit{social}, \textit{temporal} and \textit{image} signals can together help in characterizing the popular fashion boards. In particular, we study the sharing/borrowing behavior of pins and the image content characteristics of the fashion boards. We analyze the sharing behavior using social and temporal signals, and propose six novel yet simple metrics: \textit{originality score}, \textit{retention coefficients}, \textit{production coefficients}, \textit{inter-copying time}, \textit{duration of sharing} and \textit{speed coefficients}. We further study the image based content properties by extracting \textit{fashion}, \textit{color} and \textit{gender} terms embedded in the pin images. We observe significant differences across the popular (highly followed or highly ranked by the experts) and the unpopular (less followed) boards. We then use these characteristic features to early predict the popularity of a board and achieve a high correlation of $0.874$ with low RMSE value. Our key observation is that likes and repin retention coefficients are the most discriminatory factors of a board's popularity apart from the usage of various color, gender and fashion terms.
\end{abstract}

\section{Introduction}
Pinterest is an image-based online social network which has grown with unprecedented pace attaining a mark of 110 million monthly active users. It was also the fastest site to break the 10 million unique visitors mark\footnote{\url{https://techcrunch.com/2012/02/07/pinterest-monthly-uniques/}}. Although Pinterest is fairly new in the social media gamut, it is being heavily used by many big business houses like Etsy, The Gap, Allrecipes, Jettsetter, Nike, Adidas etc. to advertise their products. Further, Pinterest drives more revenue per click than Twitter or Facebook\footnote{\url{https://en.wikipedia.org/wiki/Pinterest}}. This stupendous growth makes it interesting to study Pinterest.

\subsection{Fashion industry and social media}
Social media is an amazing marketing tool for the fashion industry. Fashion brands can leverage public perception available on social media over various fashion items. The continuous feedback received by the fashion brands in the form of likes and comments on their social media posts lets them gauge and further viralize their product chain in the market. Image-based social media platforms like Pinterest, Instagram have become popular venues for fashion brand marketing and advertising. 

\subsection{Role of Pinterest in fashion industry}
Fashion is an integral part of Pinterest. All the major brands like Nike, Adidas Originals, Dolce Gabbana, Louis Vuitton etc. have their presence on the Pinterest platform. What are the characteristic features of these popular brands? Do they bear certain signatures -- social, temporal or image based -- that make them distinct from the not-so-popular ones?

In Pinterest, users save images (\textit{pins}) and categorize them on different \textit{boards}. Thus, a board is an important entity in Pinterest, and it has various influences on interest-driven pin propagation or pin sharing. Sharing is an important aspect in social media. If one likes a content, one might tend to use it in the same/modified form. On Pinterest, sharing and borrowing of pins (images) from various boards is a routine phenomena. This motivates us to consider sharing/borrowing behavior in understanding the popularity of the fashion boards. Similarly, we hypothesize that the (image) content of the post should also be a key factor determining the popularity of a fashion board.

There are multiple boards where Pinterest advertises various fashion contents. In this paper, we study the popularity of fashion boards by analyzing their originality, sharing/borrowing behavior, and the characteristics of the image content. 

\subsection{Research objectives and contributions} 
In this paper, we analyze a massive Pinterest dataset consisting of pins and boards and make the following contributions.
\begin{itemize}
 \item We investigate three major factors that can potentially characterize popularity of a fashion board: \textit{social}, \textit{temporal} (i.e., how pins are shared or borrowed across boards) and \textit{image} content characteristics (usage of fashion, color and gender terms etc. in the image) of the pins belonging to a board. 
 \item We observe that \textit{generally popular fashion boards are able to make an existing non-popular pin popular}, whereas less popular fashion boards do not exhibit this characteristic. Note that this is a very non-intuitive finding indicating a non-assortative  behavior (the `popular' pins making the non-popular pins popular) as opposed to what is usually observed in most social networks. 
 \item Another key observation is that \textit{same content in different boards achieve different levels of popularity}. If a pin has originated from a popular board, it achieves higher popularity on the originating board than the subsequent boards to which it gets shared possibly pointing to dampening of the popularity due to re-sharing. In addition, \textit{pins keep getting shared for longer durations in popular boards}.
 \item We perform extensive image analysis of the pins on the boards and extract fashion, color, and gender terms from them. Popular fashion boards have \textit{more female faces} than the unpopular ones. Further, the popular boards have a rich \textit{collection of pins in which both the gender co-appear}. We also observe significant characteristic differences between popular and unpopular boards in the usage of \textit{color and fashion words}.
 \item Our characterization further helps us to predict whether a given fashion board would become `popular' or not. In precise, we attempt to predict the popularity in terms of the future number of followers of the boards. We achieve a very high correlation coefficient of \textcolor{blue}{0.874} with very low RMSE. A post-hoc analysis of the importance of the features indicates that the likes and the repin retention coefficients are the most discriminative ones followed by the color and gender terms embedded in the image.
\end{itemize}

\subsection{Lessons for newbie fashion houses}
The insights gained from this work can highly impact the new and upcoming fashion brands. For instance, allowing for more female faces or both male and female faces together, certain color terms (\textcolor{darkwhite}{white}, black, \textcolor{blue}{blue}, \textcolor{brown}{brown}, \textcolor{brightpink}{pink} etc.) and color combinations (\textcolor{blue}{blue}-\textcolor{brightpink}{pink}, black-\textcolor{brightpink}{pink} etc.) can increase the chances of the boards getting popular. Also they could `engineer' campaigns to promote their boards in such a way that the originating boards are able to retain the `likes' and `repins' of their pins in the face of constant sharing of these pins. 

\vspace{-2mm}
\section{Related work}
\subsection{Content characteristics, sharing, and engagement}
Content sharing ensures user engagement and commitment in future~\cite{Burke:2009}. There are diverse motivations to share content on social media~\cite{Lee2012}. Apart from network structure, the content matter also play important role in sharing. Users may share useful content to appear knowledgeable or simply to help out others~\cite{wojnicki2008word}. The emotional valence behind content also drive its extent of being shared~\cite{jamali2009digging,berger2012makes}.

Though there have been various studies on diffusion, sharing, engagement in social media, very less work has been done in the domain of visual analysis of image content. Hochman et al.~\shortcite{hochman2012visualizing} show differences in local color usage, cultural production rate, varied hue intensity (blue-gray in New York vs red-yellow in Tokyo) by analyzing images from New York and Tokyo posted on Instagram. Bakhshi et al.~\shortcite{bakhshi2014faces}
study the engagement characteristic of images containing human faces. They observe that images with human faces in them, have higher chances of receiving likes and comments. Bakhshi and Gilbert~\shortcite{bakhshi2015red} study the role of color in online diffusion of pins in Pinterest. They observe that color significantly impacts the diffusion of images and adoption of content. Red, purple and pink seem to promote diffusion, while green, blue, black and yellow suppress it. 

\noindent\textit{Popularity:} There have been few studies in the domain of Fashion trend and popularity. Sanchis-ojeda et al.~\shortcite{sanchis2016detection} explore various statistical models using clients' temporal reaction to style units change for identification and quantification of linear and cyclical fashion trends. Lee et al.~\shortcite{lee2017identifying} propose a classifier for identification of fashion-related Twitter accounts whereas we focus on understanding the popular fashion boards. Hessel et al.~\shortcite{hessel2017cats} propose a relative popularity prediction framework based on content characteristics with minimal influence of other external factors like timing effects, community preferences, and social networks. Wu et al.~\shortcite{wu2017sequential} study the sequential prediction of popularity for image posts using a deep learning framework by incorporating temporal context and temporal attention into the framework.

\subsection{Fashion brand marketing}
Yamaguchi et al.~\shortcite{Yamaguchi} study the effects of visual, textual, and social factors on the popularity in a large real-world network focused on fashion. There are several studies that focus on
understanding the growing interest in social media marketing~\cite{dubois1993market,kim2012social,kim2010impacts}. Manikonda et al.~\shortcite{manikonda2015trending} study the influence of social media in various behavior of fashion brand marketing. They also analyze fashion brands' audience retention and social engagement. 

\noindent\textit{Color in affective marketing:} There are several research works that have studied the role of color in affective marketing. Most of these works focus on various kinds of advertisements, for example, the research on role of specific colors used in magazine ads~\cite{lee1989using,schindler1986color}, the efficiency of color ads compared to black and white ads~\cite{meyers1995use,sparkman1980effect}. There are also studies on understanding the effects of colors on consumer responses~\cite{bellizzi1983effects,crowley1993two}. This line of research suggests that red backgrounds elicit greater feelings of arousal than blue ones, whereas products presented against blue backgrounds are liked more than products presented against red ones~\cite{bellizzi1983effects,middlestadt1990effect}. Gray et al.~\shortcite{gray2014science} study the relationship between color coordination and `fashionableness'. They observe that maximum fashionableness is attained by selecting a color combination that is neither completely
uniform, nor completely different, i.e., fashionable outfits are those that are moderately matched, not those that are ultra-matched (``matchy-matchy'') or zero-matched (``clashing''). This balance of extremes supports Goldilocks principle regarding aesthetic preferences that seeks to balance of simplicity and complexity.

\subsection{Studies on Pinterest}
There have been several works on Pinterest. \\
\textbf{Gender Roles:} Gilbert et al.~\shortcite{gil} perform analysis focusing on the influence of gender, geography, language usage on Pinterest. They identify features of pins that could predict the activity of the board. They observe that being female on the site leads to more repins while having fewer followers. Our work draws motivation from this work where we study various social, temporal and image content factors as driving factors of popularity. Ottoni et al.~\shortcite{ottoni2013ladies} study differences in gender role in platform usage and social interaction on Pinterest. They observe that females invest more effort in reciprocating social links, are more active and generalist in content generation whereas male are more likely to be specialists and tend to describe themselves in an assertive way. Also men and women possess different interests. Chang et al.~\shortcite{chang} study users' topical specialization and homophily. \\
\textbf{User interaction and experience}: Zarro et al.~\shortcite{forte2013wedding} investigate professional and personal uses of Pinterest with interview data and observations of online activity. Zhong et al.~\shortcite{zhong} perform analysis on how borrowing behavior facilitates social interactivity and experience. Yang et al.~\shortcite{yang} study recommendation of Pinterest boards for the Twitter users. Linder et al.~\shortcite{Linder:2014} investigate social and cognitive aspects of creativity that affect the digital curation practices of everyday ideation with Pinterest users. Miller et al.~\shortcite{Miller:2015} study perception on Pinterest of the users and non-users and show that there exist differences among these two groups and how exploring Pinterest changes the non-users' experience. \\
\textbf{Diffusion, Popularity:} Zhong et al.~\shortcite{zhong2016pinning} study the impact of social ties on Pinterest. Lo et al.~\shortcite{lo2016understanding} study user activity and purchasing behavior on Pinterest for characterization of temporal user purchase intent. Lo et al.~\shortcite{Lo:2017} in another paper characterize the growth of Pinterest boards (size) and analyze how initial growth can be used to predict future growth behavior. Han et al.~\shortcite{han} study popular and viral image diffusion in Pinterest. Deeb-Swihart et al.~\shortcite{julia} study selfie presentation in everyday life on Instagram. You et al.~\shortcite{you} study various spatio-temporal patterns of Facebook photographs as well as its diffusion pattern via social ties. 

\noindent\textbf{The present work:} Our study is different from the above ones in that we analyze the popularity aspects of (fashion) boards and attempt to understand its relationship with (i) social and temporal sharing/borrowing behavior and (ii) the gender, fashion and color terms embedded in the images posted on these boards. Our analysis sheds light on strategies and mechanisms an upcoming fashion brand could adopt to make itself popular on Pinterest which can eventually enhance their overall business. 

\section{Pinterest terminologies and the dataset}
\subsection{Entities on Pinterest}
There are several entities on Pinterest. We shall provide a brief discussion on the essential functional entities of Pinterest platform.
\begin{itemize}
\item \textbf{Pin}: A pin (analogous to a post in Facebook or a tweet in Twitter) is an image which is forms the basic building block of Pinterest. Pin is a visual bookmark. Each Pin one can see on Pinterest site links back to the website it came from. The activity related to posting a pin is known as `pinning', and the user who posts a pin is known as the `pinner'. Pins can be liked and shared. Each of these pins has the following meta-data associated with it - unique pin-id, description, number of likes, number of comments, number of repins, board name, source, and content of the comments. Sharing an already existing pin is referred to as `repinning' (similar to retweet in Twitter). 
\item \textbf{Board}: A board is a user-generated collection where one saves pins. Boards can be made in secrecy or publicly. One can add collaborators to boards. Each board has a url, a name, a description (optional) and a category (optional, e.g. Art, Architecture, Celebrities, Food and Drink, Entertainment, Education, Fashion etc.). This analogy of pins and boards replicates the real-world concept of classifying images into photo albums.
\end{itemize}

\subsection{Dataset}
The dataset used in this study contains information about 0.3 million boards and their 63 million pins. We use Pinterest API v1 to crawl information about boards and pins. Board information constitutes of the following: board description, number of followers, and the creator. Pin information has the following attributes: pin description, number of likes, number of comments, number of repins, board name, and the creator. The data collection process is divided into two parts as follows
\subsubsection{Crawling of the massive dataset}
We crawl a large dataset which should be useful for doing various analysis of the fashion boards. 
\begin{itemize}
\item \textit{Initial pin collection}: We initiate the data collection process by obtaining the pin-id of 1000 pins from \url{http:// www.pinterest.com/popular/} by generating automatic scrolls. Now, each pin-id of these pins are picked and the trailing 6 digits were permuted to generate new pin-ids. About 10 million new pin-ids are generated by this process. Information of all these pins are crawled separately.
\item \textit{Massive information collection}: We extract the board-url of each of these pins from the information crawled above. About 0.3 million unique board urls are obtained. Now, for all the board-urls, board information and their individual pin's information are crawled. This results in 59 million unique pins out of a total of 63 million pins. This massive dataset is used to find out the origin of the pins.
\end{itemize}
\subsubsection{Fashion boards dataset}
We extract names of fashion boards from the following sources: i) Fashion categories on Pinterest\footnote{\url{https://www.pinterest.com/categories/}} and ii) Expert rankings from Ranker\footnote{\url{http://www.ranker.com/list/world_s-top-fashion-brands/business-and-company-info}}, Mashable\footnote{\url{http://mashable.com/2012/08/06/top-fashion-pinterest-accounts}} and Stylecaster\footnote{\url{http://stylecaster.com/fashion-pinterest-accounts}}. Finally, we could obtain information about $\sim3600$ fashion boards. We discard those boards that have very less number of pins. A total of 4 million pins are found on these 3600 boards. For each of these board urls, we crawl the detailed information about the board and its pins from Feb, 2016 till March, 2016. 

We then further categorize the 3600 boards above into following two categories of popularity: popular boards and unpopular boards. In popular boards we define two popularity classes - Highly Followed (socially ranked) Boards (HFB) and Expert Ranked Boards (ERB). We denote the unpopular ones as the Less Followed Boards (LFB). From the collection of the fashion boards, we assume the top 20\% most followed boards as HFB and the bottom 20\% as LFB. We have tried to use other percentage values also but choosing 20\% allowed us to have a sizeable data for conducting meaning experiments. The 1200 boards which we obtain from the expert rankings are noted as expert ranked boards (ERB).
\section{Characterization of fashion boards}
In this section, we shall discuss the various factors which characterize the popularity of fashion boards. There are several factors that governs the popularity of a board - the originality/novelty, sharing/borrowing behavior as well as the content on the board (the image-characteristics of the pins).
\vspace{-2mm}
\subsection{Originality}
\begin{figure}[h]
\center{
\includegraphics*[scale = 0.2,angle=0]{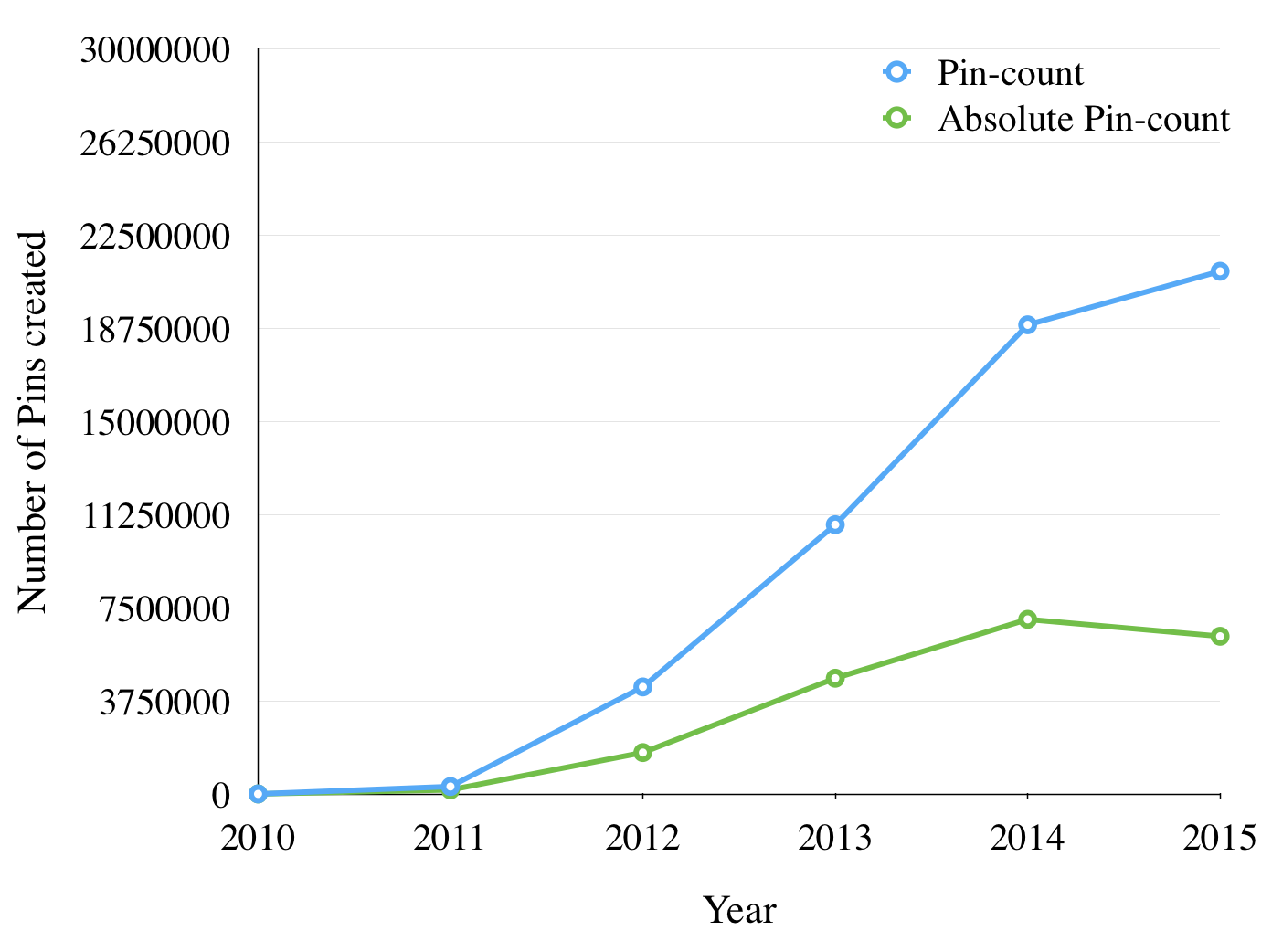}
}
\vspace{-4mm}
\caption{\label{figdataset} Evolution of unique (i.e., absolute) and total number of pins created per year.
}
\vspace{-1mm}
\end{figure}

Originality/novelty of boards is an important aspect. If one observe the creation of pins over the years (see figure~\ref{figdataset}), one can conclude that the total no. of pins are continuously growing whereas the no. of unique pins have increased only in the early few years but then started decreasing. This indicates that originality in this social media is on a decline over time due to heavy content sharing. Motivated by figure~\ref{figdataset}, we study board originality as an indicator of popularity.

\noindent\textbf{Pin originality:} Pins on a board can be classified into two types: \textit{original} pins and \textit{duplicate} pins. If a pin has originated from the board $b$, then it is called an original pin with respect to the board $b$ whereas, if a pin has not originated from the board $b$, but is a result of a copy from another board to $b$, then it is called a duplicate pin with respect to $b$.

\noindent\textbf{Board originality score:} Using the concept of pin originality, we define a measure to compute the originality score of a board. Originality score ($orig_{score}$) of a board ($b$) can be defined as the ratio of the original pins ($o_b$) on it to the total number of pins ($t_b$) on it.

$orig_{score} (b)$ = $\dfrac{o_b}{t_b}$

Originality score of a board lies in interval $[0,1]$. Boards having originality score close to 1 constitute of mainly original pins, which means that they are content generators. Boards having originality score close to 0 constitute of mainly duplicate pins, which means that they are content copiers/consumers.

\begin{figure}[h]
\center{
\includegraphics*[scale = 0.4,angle=0]{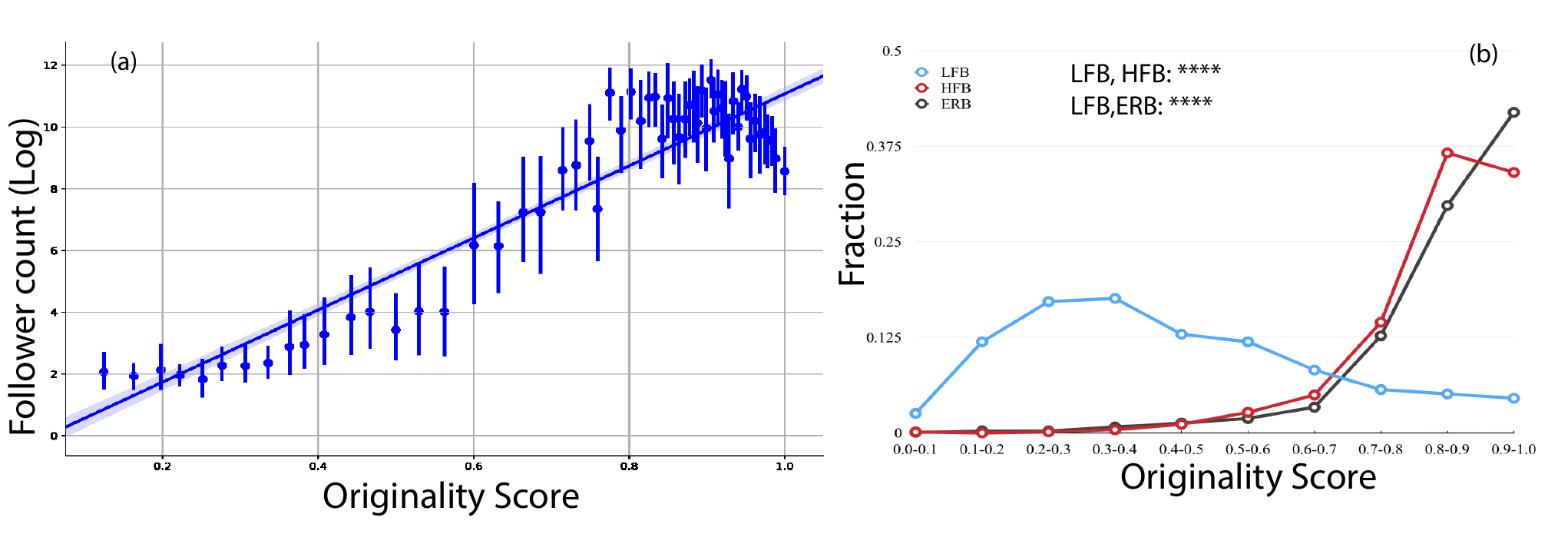}
}
\vspace{-4mm}
\caption{\label{fig_boi} (a) Relationship between originality scores and follower counts of boards. (b) Distribution of originality scores across the less followed, highly followed and expert ranked boards. The K-S test for significance among the relevant distributions are measured. ****,***,**,*, ns denote $p$-values of significance to be $<$ 0.0001, $<$ 0.001, $<$ 0.01, $<$ 0.05 and non-significant respectively. We have used the same notations for the subsequent figures wherever applicable.
}
\vspace{-1mm}
\end{figure}

In figure~\ref{fig_boi}(a), we observe that the originality scores are highly correlated with follower count. We then group the boards in less followed, highly followed and expert ranked boards and measure the originality scores in these popularity buckets. We observe that originality scores of highly followed and expert ranked boards are high, whereas that of less followed boards are on the lower side. Thus, originality of a board is an important indicator of its popularity. 

\subsubsection*{Originality of the top fashion brands}
We further study the originality scores of the boards corresponding to the top fashion brands. Toward this objective, we consider the top fashion clothing brands\footnote{\url{http://www.businessinsider.in/The-top-15-clothing-brands-millennials-love/14-Under-Armour/slideshow/51080592.cms}} and attempt to compute their originality scores. We separately collect the board information and the pins of these top fashion clothing brands. We compute the originality score of these boards and observe that they have highly original content (see table~\ref{tab:brandpop}). 

We then attempt to find the extent of correlation between the originality scores of these boards with their popularity (in terms of the number of followers). The Spearman's rank correlation comes out to be $0.41$. This establishes that there is a strong positive correlation between originality and popularity of the top fashion brands.

\begin{table}[h]
\center{
\caption{\label{tab:brandpop} Originality scores of top fashion brands}
\resizebox{6cm}{!}{
\begin{tabular}{|c|c|c|}
\hline
Rank & Brand Name        & Originality Score \\ \hline
1    & Nike              & 0.893333494064    \\ \hline
2    & Target            & 0.99560813134     \\ \hline
3    & Adidas            & 1.0               \\ \hline
4    & Macy's            & 0.98539077817     \\ \hline
5    & JCPenney          & 0.992822203076    \\ \hline
6    & Converse          & 0.877229331742    \\ \hline
7    & Van's             & 0.990441463251    \\ \hline
8    & Ralph Lauren      & 0.996835443038    \\ \hline
9    & Forever 21        & 0.953847264232    \\ \hline
10   & Victoria's Secret & 0.988732432462    \\ \hline
11   & Levi's            & 0.944628465324    \\ \hline
12   & Chanel            & 0.876407556394    \\ \hline
13   & Under Armour      & 0.897653427232    \\ \hline
14   & Aeropostale       & 0.916352963232    \\ \hline
\end{tabular}}
}
\end{table}
\vspace{-2mm}
\subsection{Sharing/borrowing behavior}
Sharing/borrowing of pins are very common on the Pinterest platform. We introduce {\em board retention coefficients} and {\em board production coefficients} based on the sharing/borrowing behavior dynamics of the pins on a board. On Pinterest, the `social behavior' of a pin can be measured based on three factors: the {\em number of likes}, the {\em number of repins} and the {\em number of comments} generated by the pin. We however observe that commenting is not practiced extensively in this platform. Hence, we only take number of $likes$ and $repins$ generated by a pin. The two coefficients we define next roughly correspond to the direction and magnitude of flow of information from one board to another. Each of them is independently able to portray meaningful information about sharing.

\subsubsection{\textbf{Retention coefficients}}
\textit{Board retention coefficients} are a novel set of measures concerning the $like$/$repin$ `retention' capabilities of a board. It addresses the question that - {\em how many likes/repins shall a board be able to retain if other boards copy content from it}. We calculate the \textit{likes retention coefficient} using the  algorithm~\ref{algo_lrc}. Similarly, we also compute the \textit{repins retention coefficient}.

\begin{algorithm}[h]
\caption{\label{algo_lrc}Calculation of {\em likes retention coefficient}.}
\begin{algorithmic}
\STATE temp $\leftarrow$ [ ]
\FOR{each original pin $p$ on board $b$}
\STATE   $temp.append(\dfrac{1 + likes \: of \: p \: on \: b}{1 + avg. \: likes \: of \:p \:on \:other \: boards})$
\ENDFOR
\STATE likes retention coefficient of board $b$ = average(temp)
\end{algorithmic}
\end{algorithm}

Both the retention coefficient values lie in the interval $(1,\infty)$. If a board $b$ has a higher likes (repins) retention coefficient, then the subsequent boards that copy pins from this board shall be able to garner less likes (repins) than the board $b$.

\begin{figure}[h]
\center{
\includegraphics[scale = 0.4,angle=0]{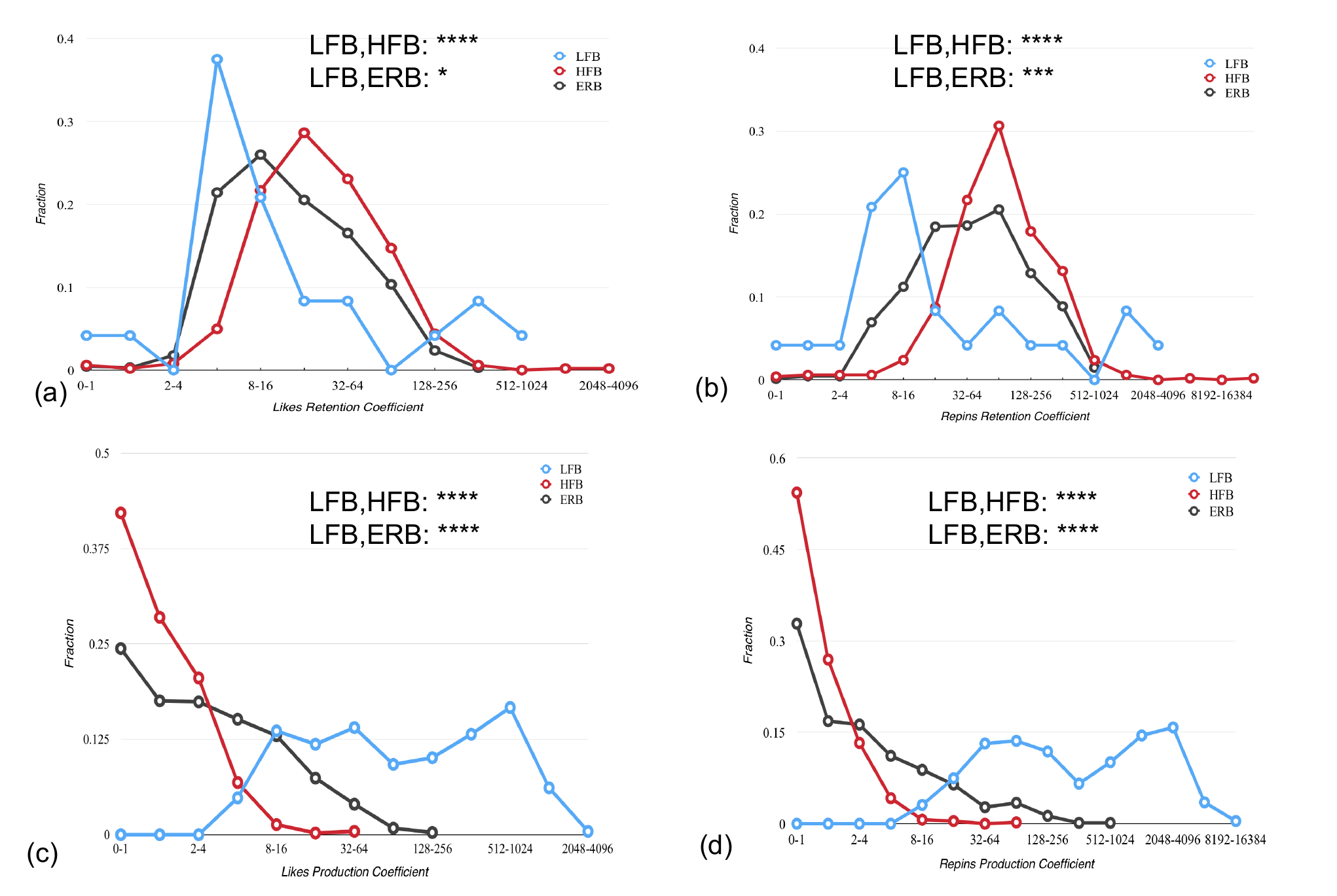}
}
\vspace{-4mm}
\caption{\label{fig_coeff} Distribution of a) likes retention coefficient b) repins retention coefficient (c) likes production coefficient b) repins production coefficient for less followed, highly followed and expert ranked boards.
}
\vspace{-1mm}
\end{figure}

A significant fraction of highly followed and expert ranked boards have higher retention coefficients compared to the less followed boards (see figure~\ref{fig_coeff}(a) and (b)). Thus, the original pins on highly followed and expert ranked boards are significantly more liked/repinned among their respective duplicate (shared) pins, whereas the original pins on less followed boards are not popular among their duplicate pins. Hence, the likes/repins of the content on highly followed and expert ranked boards do not decline even after they are duplicated through copying.

\vspace{-2mm}
\subsubsection{\textbf{Production coefficients}}
\textit{Board production coefficients} are a novel set of measures that capture the like/repin production capacities of a board. It addresses the question that - {\em how many likes/repins shall other boards gain if they copy content from a board}. We compute the \textit{likes production coefficient} using the algorithm~\ref{algo_lpc}. Similarly, we also compute \textit{repins production coefficient}.

\begin{algorithm}
\caption{\label{algo_lpc} Calculation of {\em likes production coefficient}.}
\begin{algorithmic}
\STATE temp $\leftarrow$ []
\FOR{each duplicate pin $p$ on board $b$}
\STATE  $temp.append(\dfrac{1 + likes\: of\: p \:on \:its \:original\: board}{1 + likes \: of \: p \:on\: b})$
\ENDFOR
\STATE likes production coefficient of board $b$ = average(temp)
\end{algorithmic}
\end{algorithm}

Both the production coefficients lie in the interval $(1,\infty)$. If a board $b$ has a lower likes (repins) production coefficient, then the duplicate pins on $b$ shall garner more likes (repins) compared to that on their board of origin.


A vast majority of highly followed and expert ranked boards have lower production coefficients than the less followed boards (see figure~\ref{fig_coeff} (c) and (d)). Thus, the duplicate pins on highly followed and expert ranked boards generate more number of likes/repins compared to that generated on their corresponding boards of origin. On the other hand, duplicate pins on less followed boards generate less likes/repins compared to that in their corresponding boards of origin. Hence, highly followed boards and expert ranked boards are able to make an existing pin more liked/repinned by copying it.
\vspace{-2mm}
\subsection{Temporal dynamics of sharing/borrowing}
In this section, we introduce two measures based on the temporal aspects of sharing: {\em inter-copying time} and {\em duration of sharing}. In addition, we also define \textit{speed coefficients} that indicate the speed at which likes/repins are gained.
\begin{figure*}[!ht]
\center{
\includegraphics[scale = 0.4,angle=0]{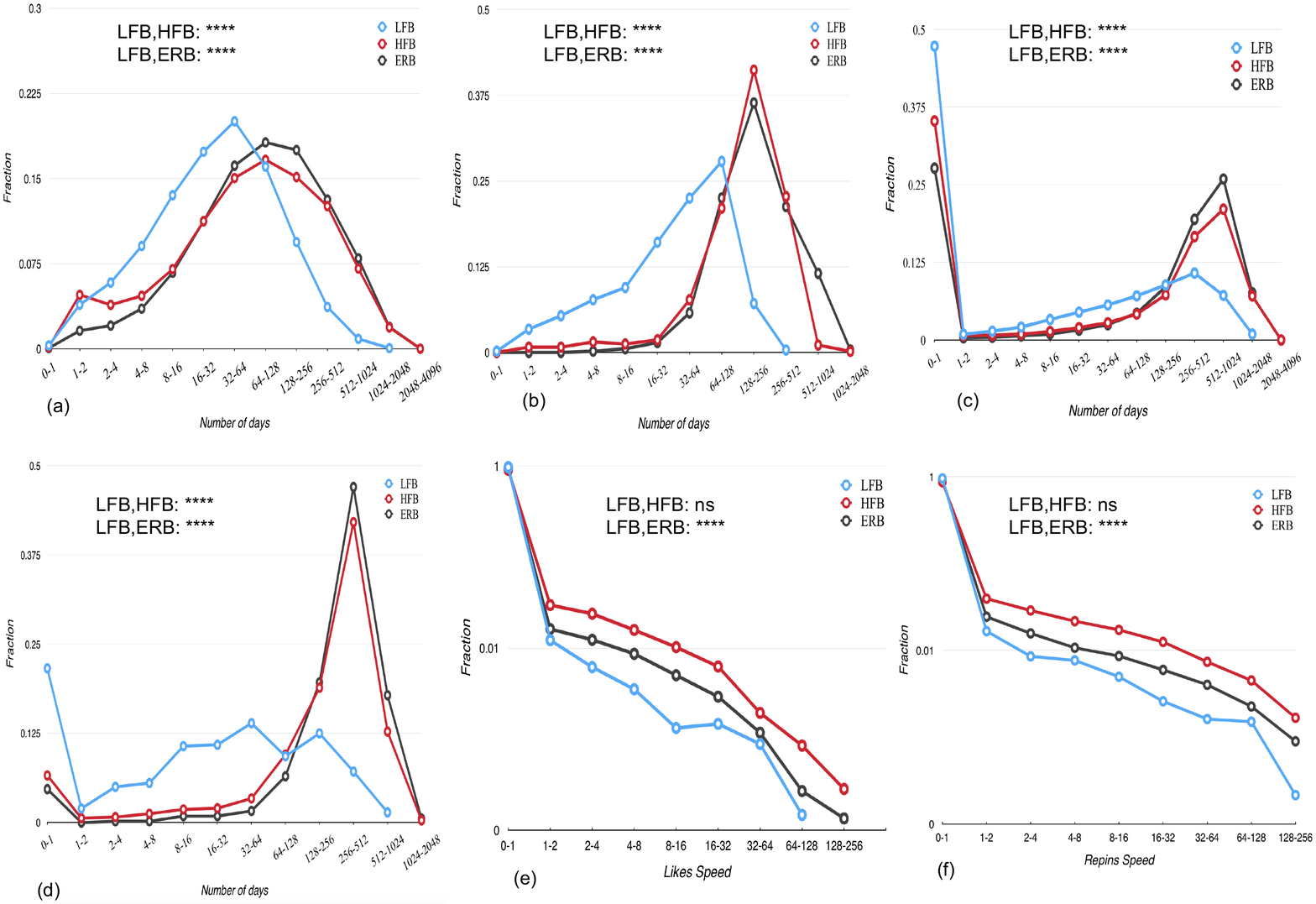}
}
\vspace{-4mm}
\caption{\label{fig_temporal} Distribution of a) inter-copying time for pins b) inter-copying time for boards c) duration of sharing for pins d) duration of sharing for boards e) likes speed coefficient f) repins speed coefficient for pins in less followed, highly followed and expert ranked boards.
}
\vspace{-1mm}
\end{figure*}
{\em Inter-copying time} is a measure defined for the original pins.  This is expressed as the average time-gaps between instances of sharing of an original pin on the subsequent boards. For an original pin $p$, we compute {\em ICT} as explained in algorithm~\ref{algo_ict}. We now average the value of {\em ICT}s for all the original pins on board $b$, and call this as {\em inter-copying time} of board $b$.

A significant number of pins belonging to highly followed and expert ranked boards have a higher value of inter-copying time than pins on less followed boards (see figure~\ref{fig_temporal}(a) and (b)). This shows that pins on less followed boards have smaller time-gaps between consecutive shares as compared to pins on highly followed and expert ranked boards.

\begin{algorithm}[h]
\caption{\label{algo_ict} Calculation of {\em inter-copying time} for an original pin $p$.}
\begin{algorithmic}
\STATE $temp$ $\leftarrow$ []
\FOR{each duplicate pin $p^{'}$ generated from pin $p$}
\STATE $temp$.append(time-stamp of $p^{'}$)
\ENDFOR
\STATE sort $temp$ in non-decreasing order
\FOR{each $i$ in range(0, len($temp$))}
\IF{$i$ == 0}
\STATE $temp[i]$ $\leftarrow$ 0
\ELSE
\STATE $temp[i]$ $\leftarrow$ $temp[i]$ - $temp[i-1]$
\ENDIF
\ENDFOR
\STATE $ICT$ for pin $p$ $\leftarrow$ $average(temp)$
\end{algorithmic}
\end{algorithm}

\subsubsection{\textbf{Duration of sharing (DoS)}}
Similar to {\em inter-copying time}, {\em duration of sharing} is also defined for original pins. It can be interpreted as the life-cycle of sharing of a pin. For an original pin $p$, we compute {\em DoS} as explained in algorithm~\ref{algo_dos}. We now average the value of {\em DoS}s for all original pins on board $b$, and call this as {\em duration of sharing} of board $b$.

\begin{algorithm}[h]
\caption{\label{algo_dos} Calculation of {\em duration of sharing} for an original pin $p$.}
\begin{algorithmic}
\STATE $temp$ $\leftarrow$ []
\FOR{each duplicate pin $p^{'}$ generated from pin $p$}
\STATE $temp$.append(time-stamp of $p^{'}$)
\ENDFOR
\STATE sort $temp$ in non-decreasing order
\STATE $DoS$ for pin $p$ $\leftarrow$ $temp[len($temp$)-1] - temp[0]$
\end{algorithmic}
\end{algorithm}

A large fraction of pins on highly followed boards and expert ranked boards have high duration of sharing compared to the less followed boards (see figure~\ref{fig_temporal}(c) and (d)). Hence, a pin on highly followed and expert ranked boards is likely going to have a longer life span than a pin on the less followed boards.


\subsubsection{\textbf{Speed coefficients}}
In this section, we attempt to combine the likes/repins on a board with its temporal characteristics. Toward this objective, we define likes and repins speed coefficients as follows.


we compute {\em likes speed coefficient} as explained in algorithm~\ref{algo_lsc}. We then average the value of {\em likes speed coefficient} for all the original pins on a board $b$, and call this as the {\em likes speed coefficient} of board $b$. We similarly calculate {\em repins speed coefficient}.

\begin{algorithm}[h]
\caption{\label{algo_lsc} Calculation of {\em likes speed coefficient} for an original pin $p$.}
\begin{algorithmic}
\STATE $likes$ $\leftarrow$ []
\FOR{each duplicate pin $p^{'}$ generated from pin $p$}
\STATE $likes$.append(number of likes on $p^{'}$)
\ENDFOR
\STATE {\em likes speed coefficient} of $p$ $\leftarrow$ $\dfrac{sum(likes)}{DoS(p)}$
\end{algorithmic}
\end{algorithm}

Speed coefficients are greater for highly followed and expert ranked boards compared to the less followed boards (see figure~\ref{fig_temporal}(e) and (f))). Thus, original pins on highly followed and expert ranked boards gain popularity much more quickly than the original pins on less followed boards.


\subsection{Image-based content analysis}
Pinterest being an image sharing social media, the characteristics of image (pins) also should have an impact on their popularity. In this section, we analyze the content characteristics of images (pins) on the various boards. Toward this objective, we use densecap~\cite{densecap}, an image captioning tool that extracts salient regions from an image and describes them in natural language (English). We perform tokenization, stemming, lemmatization and stop-words removal of these generated `dense' captions to obtain key tokens/phrases demonstrating the image. We further group these tokens into three key types: \textit{gender terms}, \textit{fashion terms} and \textit{color terms}. Gender terms which we analyze are {\em male} and {\em female}. We obtain an exhaustive listing of fashion terms from Myvocabulary\footnote{\url{https://myvocabulary.com/word-list/fashion-and-clothing-vocabulary/}}. A universal set of all color terms are available in Wikipedia\footnote{\url{https://en.wikipedia.org/wiki/Lists_of_colors}}. The fashion and color terms which we find both in the above respective listings and our data-set are shown in table~\ref{tab_fashion_items} and ~\ref{tab_colors}.
\begin{table}[h]
\center{
\vspace{-2mm}
\caption{Fashion terms in the dataset.
\vspace{-2mm}}
\label{tab_fashion_items}
\resizebox{6cm}{!}{
\begin{tabular}{|c|c|c|c|c|}
\hline
clothes  & tshirt   & skin       & shirts & jacket  \\ \hline
feathers & pillows  & sunglasses & buttons & shoe \\ \hline
suit     & curtains & skirt      & leather &  pants \\ \hline
trouser  & striped  & shorts     & strap & dress   \\ \hline
jeans    & pillow   & necklace   & umbrella & bag  \\ \hline
\end{tabular}
}}
\end{table}

\begin{table}[h]
\center{
\vspace{-2mm}
\caption{Color terms in the dataset.
\vspace{-2mm}}
\label{tab_colors}
\resizebox{7cm}{!}{
\begin{tabular}{|c|c|c|c|c|c|c|}
\hline
black & \textcolor{blue}{blue} & \textcolor{darkwhite}{white} & \textcolor{brown}{brown} & \textcolor{green}{green} & \textcolor{purple}{purple} & \textcolor{red}{red}\\ \hline
\textcolor{yellow}{yellow} & \textcolor{gray}{grey} & \textcolor{metal}{metal} & \textcolor{wooden}{wooden} & \textcolor{brightpink}{pink} & \textcolor{silver}{silver} & \\ \hline
\end{tabular}
}}
\end{table}

\subsubsection{\textbf{Gender term analysis}}
We analyze the occurrences of both the genders in the pins across all the three categories of boards. In table~\ref{tab_img1}, we report the occurrences of each gender. The cell corresponding to {\em Male} and {\em Less Followed Boards} has a value of 0.45. This means that 0.45 fraction of pins belonging to less followed boards have male faces on them. We observe that the number of {\em female} faces on highly followed boards and expert ranked boards is high, whereas they are significantly lower (\textbf{$\sim$ 20\%} lower) in less followed boards. This shows that having more female faces on a board could increase the popularity of a board. Further, if a board has more female faces, it has more chances of being listed in expert ranked boards.
\begin{table}[h]
\center{
\caption{Fraction of pins on boards with various gender combinations.}
\label{tab_img1}
\resizebox{5cm}{!}{
\begin{tabular}{|c|c|c|c|}
\hline
  Gender & LFB & HFB & ERB\\ \hline
Male only   & 0.45                 & 0.48                   & 0.49                 \\ \hline
Female only  & 0.49                 & \cellcolor{pink}\textbf{0.57}                   & \cellcolor{pink}\textbf{0.61}                 \\ \hline
Male-Female & 0.26                 & \cellcolor{pink}\textbf{0.37}                   & \cellcolor{pink}\textbf{0.35}                 \\ \hline
\end{tabular}
}}
\end{table}
Another very interesting observation is that the more popular boards have a higher fraction of pins containing both male and female faces together on a single pin compared to the less followed boards. 

In summary, highly followed boards and expert ranked boards have more female faces than the less followed boards. Moreover, the boards in the former category have a richer collection of pins that together feature faces of the genders.

\subsubsection{\textbf{Fashion term analysis}}
We analyze the occurrences of various fashion terms in table~\ref{tab_fashion_items} in the imagery of the pins across all the three categories of boards. We compute number of occurrences of the fashion terms from table~\ref{tab_fashion_items} appearing in the pins across the boards. We use the torso of this frequency distribution to extract the most discerning fashion terms. We choose the top 10 fashion terms from the torso to compute results in table~\ref{tab_img3_0}. Each cell $(x, y)$ in the table represents the fraction of pins belonging to a particular popularity class `y' having the fashion term `x'. Therefore, the cell value of 0.318 corresponding to {\em shirt} and {\em Less Followed Boards} means that 0.318 fraction of pins belonging to less followed boards have the fashion term {\em shirt} in them. We observe that 20\% of fashion terms have equal distributions in highly followed boards and expert ranked boards whereas their distributions in the less followed boards are quite different.
\begin{table}[h]
\centering
\vspace{-2mm}
\caption{Fraction of pins having top 10 fashion terms which are obtained from the torso of the frequency distribution of all the fashion terms in table~\ref{tab_fashion_items}. The boldface values indicate similar distribution among $ERB$ and $HFB$ but a different distribution in LFB. We adopted this convention in the subsequent tables also.}
\vspace{-2mm}
\label{tab_img3_0}
\resizebox{6cm}{!}{
\begin{tabular}{|c|c|c|c|}
\hline
Fashion Terms & \begin{tabular}[c]{@{}c@{}}LFB\end{tabular} & \begin{tabular}[c]{@{}c@{}}HFB\end{tabular} & \begin{tabular}[c]{@{}c@{}}ERB\end{tabular} \\ \hline
shirt & 0.318 & 0.389 & 0.330 \\ \hline
\cellcolor{pink}\textbf{bag} & \cellcolor{pink}\textbf{0.221} & \cellcolor{pink}\textbf{0.291} & \cellcolor{pink}\textbf{0.312} \\ \hline
dress & 0.221 & 0.179 & 0.225 \\ \hline
pants & 0.210 & 0.154 & 0.216 \\ \hline
shoe & 0.187 & 0.231 & 0.201 \\ \hline
jacket & 0.159 & 0.148 & 0.139 \\ \hline
umbrella & 0.158 & 0.139 & 0.146 \\ \hline
necklace & 0.147 & 0.134 & 0.156 \\ \hline
pillow & 0.143 & 0.131 & 0.153 \\ \hline
\cellcolor{pink}\textbf{jeans} & \cellcolor{pink}\textbf{0.092} & \cellcolor{pink}\textbf{0.127} & \cellcolor{pink}\textbf{0.152} \\ \hline
\end{tabular}}
\end{table}

We further study the co-occurrences of fashion terms in more detail. In table~\ref{tab_img3}, we report the number of occurrences of two co-occurring fashion terms in a pin. Each cell $(x\textrm{-}z, y)$ in the table represents the fraction of pins belonging to a particular popularity class `y' having the co-occurring fashion term `x-z'. Thus, the value of 0.0964 corresponding to {\em jacket-trouser} and {\em Less Followed Boards} means that 0.0964 fraction of pins belonging to {\em Less Followed Boards} have both {\em jacket} and {\em trouser} together in them. We observe that \textbf{70\%} co-occurring bi-terms have almost similar distributions in highly followed boards and expert ranked boards, whereas their distributions in the less followed boards are very different from the other two. 

\begin{table}[h]
\center{
\vspace{-2mm}
\caption{Fractions of pins having top 10 co-occurring bi-terms, which are obtained from the torso of the distribution of the fashion terms in table~\ref{tab_fashion_items}.
\vspace{-2mm}}
\label{tab_img3}
\resizebox{7.5cm}{!}{
\begin{tabular}{|c|c|c|c|}
\hline
 Bi-terms & \begin{tabular}[c]{@{}c@{}}LFB\end{tabular} & \begin{tabular}[c]{@{}c@{}}HFB\end{tabular} & \begin{tabular}[c]{@{}c@{}}ERB\end{tabular} \\ \hline
\cellcolor{pink}\textbf{jacket-trouser} & \cellcolor{pink}\textbf{0.0964} & \cellcolor{pink}\textbf{0.1276} & \cellcolor{pink}\textbf{0.1198} \\ \hline
\cellcolor{pink}\textbf{bag-umbrella} & \cellcolor{pink}\textbf{0.0753} & \cellcolor{pink}\textbf{0.1243} & \cellcolor{pink}\textbf{0.1134} \\ \hline
\cellcolor{pink}\textbf{bag-striped} & \cellcolor{pink}\textbf{0.0683} & \cellcolor{pink}\textbf{0.1223} & \textbf{0.1143} \\ \hline
necklace-strap & 0.0879 & 0.1124 & 0.0953 \\ \hline
\cellcolor{pink}\textbf{jeans-shoe} & \cellcolor{pink}\textbf{0.0762} & \cellcolor{pink}\textbf{0.1057} & \cellcolor{pink}\textbf{0.1049} \\ \hline
bag-shorts & 0.1032 & 0.1242 & 0.0643 \\ \hline
\cellcolor{pink}\textbf{bag-trouser} & \cellcolor{pink}\textbf{0.0923} & \cellcolor{pink}\textbf{0.1243} & \cellcolor{pink}\textbf{0.1142} \\ \hline
\cellcolor{pink}\textbf{leather-strap} & \cellcolor{pink}\textbf{0.0923} & \cellcolor{pink}\textbf{0.1214} & \cellcolor{pink}\textbf{0.1203} \\ \hline
\cellcolor{pink}\textbf{dress-umbrella} & \cellcolor{pink}\textbf{0.0812} & \cellcolor{pink}\textbf{0.1143} & \cellcolor{pink}\textbf{0.1043} \\ \hline
dress-skirt & 0.1023 & 0.0854 & 0.1053 \\ \hline
\end{tabular}
}}
\end{table}

We also consider the co-occurring tri-terms (three fashion terms together). We observe similar discriminating results for \textbf{60\%} tri-terms (see table~\ref{tab_img4}). The discrimination becomes more prominent when we use the co-occurring bi-terms and tri-terms. We thus conclude that the collection of fashion terms together used in pins affect the popularity of their boards. Hence, a popularity seeking board can host images having a particular collection of fashion terms from the above analysis.
\begin{table}[h]
\center{
\vspace{-2mm}
\caption{Fraction of pins having top 10 co-occurring tri-terms, which are obtained from the torso of the distribution of the fashion terms in table~\ref{tab_fashion_items}.
\vspace{-2mm}}
\label{tab_img4}
\begin{tabular}{|c|c|c|c|}
\hline
Tri-terms & \begin{tabular}[c]{@{}c@{}}LFB\end{tabular} & \begin{tabular}[c]{@{}c@{}}HFB\end{tabular} & \begin{tabular}[c]{@{}c@{}}ERB\end{tabular} \\ \hline
\cellcolor{pink}\textbf{leather-pillow-shirt} & \cellcolor{pink}\textbf{0.1032} & \cellcolor{pink}\textbf{0.1343} & \cellcolor{pink}\textbf{0.1763} \\ \hline
\cellcolor{pink}\textbf{pants-strap-trouser} & \cellcolor{pink}\textbf{0.0913} & \cellcolor{pink}\textbf{0.1132} & \cellcolor{pink}\textbf{0.1298} \\ \hline
\cellcolor{pink}\textbf{pants-shoe-skirt} & \cellcolor{pink}\textbf{0.0613} & \cellcolor{pink}\textbf{0.1232} & \cellcolor{pink}\textbf{0.1265} \\ \hline
jeans-leather-pants & 0.1265 & 0.0942 & 0.1135 \\ \hline
jeans-shirt-shorts & 0.1175 & 0.0823 & 0.0732 \\ \hline
bag-necklace-skirt & 0.0786 & 0.0974 & 0.1296 \\ \hline
dress-shirt-sunglasses & 0.0874 & 0.0925 & 0.1145 \\ \hline
\cellcolor{pink}\textbf{bag-pants-trouser} & \cellcolor{pink}\textbf{0.0874} & \cellcolor{pink}\textbf{0.1134} & \cellcolor{pink}\textbf{0.1341} \\ \hline
\cellcolor{pink}\textbf{bag-shirt-umbrella} & \cellcolor{pink}\textbf{0.0112} & \cellcolor{pink}\textbf{0.1324} & \cellcolor{pink}\textbf{0.1142} \\ \hline
\cellcolor{pink}\textbf{dress-pants-shorts} & \cellcolor{pink}\textbf{0.0931} & \cellcolor{pink}\textbf{0.1121} & \cellcolor{pink}\textbf{0.1321} \\ \hline
\end{tabular}}
\end{table}

\subsubsection{\textbf{Color term analysis}}
Colors are a very important factor in fashion~\cite{bakhshi2015red}. In this section, we analyze the occurrence of color terms from table~\ref{tab_colors} appearing in the pins across all three categories of boards. We compute the number of occurrences of each color term, co-occurring bi-terms generated from the colors in table~\ref{tab_colors}. We choose the top 10 color terms (once again, from the torso of the frequency distribution) in table~\ref{tab_img5}. We observe that 30\% color terms have similar distributions in the highly followed boards and expert ranked boards, whereas their distributions in the less followed boards are different from the other two. \textcolor{darkwhite}{White}, black and \textcolor{blue}{blue} are found to be the three mostly used color terms whereas \textcolor{purple}{purple} is the least favored one.

In table~\ref{tab_img6}, we show the occurrence distribution of the top 10 most co-occurring color bi-terms from the torso of the frequency distribution. Once again, we obtain a better discrimination while using bi-terms over single term occurrences (\textbf{40\%} over \textbf{30\%}). Black-\textcolor{yellow}{yellow} and \textcolor{blue}{blue}-\textcolor{yellow}{yellow} are the most co-occurring color terms for the less followed boards whereas black-\textcolor{brightpink}{pink} and \textcolor{brightpink}{pink}-\textcolor{red}{red} are the most dominating color terms occurring together for the highly followed boards in the torso region. For the expert ranked boards, \textcolor{blue}{blue}-\textcolor{brightpink}{pink} and \textcolor{brightpink}{pink}-\textcolor{red}{red} are the most used color terms whereas \textcolor{blue}{blue}-\textcolor{darkwhite}{white}, black-\textcolor{darkwhite}{white}, \textcolor{blue}{blue}-black are the most dominant color combinations in all the three categories of boards when we consider the whole distribution. Therefore, we observe that the color composition of images (pins) affect the popularity of their boards. Hence, a popularity seeking board can host images having a particular color composition from the above analysis.

\begin{table}[h]
\centering
\vspace{-2mm}
\caption{Fractions of pins having top 10 color terms which are obtained from the torso of the frequency distribution of all colors terms mentioned in table~\ref{tab_colors}.}
\vspace{-2mm}
\label{tab_img5}
\begin{tabular}{|c|c|c|c|}
\hline
Color & \begin{tabular}[c]{@{}c@{}}LFB\end{tabular} & \begin{tabular}[c]{@{}c@{}}HFB\end{tabular} & \begin{tabular}[c]{@{}c@{}}ERB\end{tabular} \\ \hline
\textcolor{darkwhite}{white} & \textcolor{darkwhite}{0.638} & \textcolor{darkwhite}{0.664} & \textcolor{darkwhite}{0.696} \\ \hline
black & 0.555 & 0.581 & 0.538 \\ \hline
\textcolor{blue}{blue} & \textcolor{blue}{0.543} & \textcolor{blue}{0.566} & \textcolor{blue}{0.492} \\ \hline
\textcolor{brown}{\textbf{brown}} & \textcolor{brown}{\textbf{0.497}} & \textcolor{brown}{\textbf{0.379}} & \textcolor{brown}{\textbf{0.389}} \\ \hline
\textcolor{red}{\textbf{red}} & \textcolor{red}{\textbf{0.346}} & \textcolor{red}{\textbf{0.244}} & \textcolor{red}{\textbf{0.262}} \\ \hline
\textcolor{wooden}{\textbf{wooden}} & \textcolor{wooden}{\textbf{0.338}} & \textcolor{wooden}{\textbf{0.218}} & \textcolor{wooden}{\textbf{0.236}} \\ \hline
\textcolor{green}{green} & \textcolor{green}{0.224} & \textcolor{green}{0.221} & \textcolor{green}{0.233} \\ \hline
\textcolor{metal}{metal} & \textcolor{metal}{0.256} & \textcolor{metal}{0.205} & \textcolor{metal}{0.198} \\ \hline
\textcolor{brightpink}{pink} & \textcolor{brightpink}{0.122} & \textcolor{brightpink}{0.098} & \textcolor{brightpink}{0.086} \\ \hline
\textcolor{purple}{purple} & \textcolor{purple}{0.012} & \textcolor{purple}{0.008} & \textcolor{purple}{0.012} \\ \hline
\end{tabular}
\end{table}

\begin{table}[h]
\centering
\vspace{-2mm}
\caption{Fractions of pins having top 10 co-occurring bi-terms, which are obtained from the torso of the distribution of all possible color bi-terms from table~\ref{tab_colors}.}
\vspace{-2mm}
\label{tab_img6}
\resizebox{7cm}{!}{
\begin{tabular}{|c|c|c|c|}
\hline
Color bi-terms & \begin{tabular}[c]{@{}c@{}}LFB\end{tabular} & \begin{tabular}[c]{@{}c@{}}HFB\end{tabular} & \begin{tabular}[c]{@{}c@{}}ERB\end{tabular} \\ \hline
black-\textcolor{yellow}{yellow} & 0.1043 & 0.0745 & 0.0943 \\ \hline
\textbf{\textcolor{blue}{blue}-\textcolor{brightpink}{pink}} & \textbf{0.0744} & \textbf{0.0935} & \textbf{0.1064} \\ \hline
\textbf{black-\textcolor{brightpink}{pink}} & \textbf{0.0824} & \textbf{0.1053} & \textbf{0.1034} \\ \hline
\textcolor{metal}{metal}-\textcolor{red}{red} & 0.0743 & 0.0723 & 0.1053 \\ \hline
\textcolor{blue}{blue}-\textcolor{yellow}{yellow} & 0.1024 & 0.0923 & 0.0814 \\ \hline
\textcolor{blue}{blue}-\textcolor{silver}{silver} & 0.0908 & 0.0956 & 0.0932 \\ \hline
\textbf{\textcolor{brightpink}{pink}-\textcolor{red}{red}} & \textbf{0.0824} & \textbf{0.1025} & \textbf{0.1057} \\ \hline
\textcolor{metal}{metal}-\textcolor{silver}{silver} & 0.0823 & 0.0675 & 0.0723 \\ \hline
\textbf{\textcolor{red}{red}-\textcolor{yellow}{yellow}} & \textbf{0.0923} & \textbf{0.0814} & \textbf{0.0774} \\ \hline
\textcolor{gray}{grey}-\textcolor{darkwhite}{white} & 0.0452 & 0.0423 & 0.0424 \\ \hline
\end{tabular}}
\end{table}

\subsubsection{\textbf{Gender infused fashion analysis}}
In this section, we shall study gender based usage of fashion and color terms across the three board categories. In table~\ref{tab_img7}, we show the gender based usage of the fashion terms (bi-terms). Each cell $(x, y)$ in the table represents the fraction of pins belonging to a particular popularity class `y' having the gender (`g') based fashion bi-term `a-b'. Here `x' corresponds to `g-a-b'. For example, the cell value corresponding to {\em man-bag-jeans} and {\em Less Followed Boards} means that 0.1375 fraction of pins belonging to less followed boards have male, bag and jeans in them. We observe that \textbf{50\%} of combinations have almost equal distributions in the highly followed boards and expert ranked boards, whereas their distributions in the less followed boards are very different from the other two. Such differences in the combination of gender and fashion terms affect the popularity of boards. It is seen that some combinations increase the popularity of boards, whereas rest decrease it. Another interesting observation we obtain from this analysis is that all the 40\% combinations which have {\em woman} in them correspond to more popular boards.
\begin{table}[h]
\centering
\vspace{-2mm}
\caption{Fractions of pins having top 10 gender-based fashion bi-terms.}
\vspace{-2mm}
\label{tab_img7}
\resizebox{7cm}{!}{
\begin{tabular}{|c|c|c|c|}
\hline
Gender and Fashion Bigrams & LFB & HFB & ERB\\ \hline
man-bag-jeans & 0.1375 & 0.1323 & 0.1250 \\ \hline
\cellcolor{pink}\textbf{man-dress-shorts} & \cellcolor{pink}\textbf{0.1175} & \cellcolor{pink}\textbf{0.1442} & \cellcolor{pink}\textbf{0.1525} \\ \hline
\cellcolor{pink}\textbf{woman-bag-jeans} & \cellcolor{pink}\textbf{0.1275} & \cellcolor{pink}\textbf{0.1567} & \cellcolor{pink}\textbf{0.1489} \\ \hline
\cellcolor{pink}\textbf{woman-bag-strap} & \cellcolor{pink}\textbf{0.1325} & \cellcolor{pink}\textbf{0.1578} & \cellcolor{pink}\textbf{0.1652} \\ \hline
woman-shirt-striped & 0.1450 & 0.1682 & 0.1575 \\ \hline
\cellcolor{pink}\textbf{man-bag-shoe} & \cellcolor{pink}\textbf{0.1675} & \cellcolor{pink}\textbf{0.1324} & \cellcolor{pink}\textbf{0.1434} \\ \hline
woman-bag-shoe & 0.1424 & 0.1550 & 0.1523 \\ \hline
\cellcolor{pink}\textbf{woman-shirt-skirt} & \cellcolor{pink}\textbf{0.1375} & \cellcolor{pink}\textbf{0.1576} & \cellcolor{pink}\textbf{0.1503} \\ \hline
woman-necklace-pants & 0.1324 & 0.1425 & 0.1232 \\ \hline
man-shirts-shorts & 0.1453 & 0.1486 & 0.1502 \\ \hline
\end{tabular}
}
\end{table}

\subsubsection{\textbf{Gender infused color analysis}}
In table~\ref{tab_img8}, we show the distribution of top five gender-based color bi-terms among pins across the three board categories. We observe that $\sim$60\% combinations have equal distributions in highly followed boards and expert ranked boards, whereas their distributions in less followed boards are different from the other two. Though black-\textcolor{metal}{metal} is the dominant color combination for female in both less and highly followed boards, \textcolor{darkwhite}{white}-\textcolor{metal}{metal} is the prominent color combination in expert ranked boards. In general, \textcolor{metal}{metal} colors seem to go very well with women.
\begin{table}[h]
\vspace{-2mm}
\centering
\caption{Fractions of pins having top five gender-based co-occurring color bi-terms.}
\label{tab_img8}
\vspace{-2mm}
\resizebox{7cm}{!}{
\begin{tabular}{|c|c|c|c|}
\hline
Gender-Color Trigrams & \begin{tabular}[c]{@{}c@{}}LFB\end{tabular} & \begin{tabular}[c]{@{}c@{}}HFB\end{tabular} & \begin{tabular}[c]{@{}c@{}}ERB\end{tabular} \\ \hline
woman-metal-white & 0.2853 & 0.2753 & 0.2895 \\ \hline
\textbf{woman-\textcolor{brightpink}{pink}-\textcolor{darkwhite}{white}} & \textbf{0.2514} & \textbf{0.2657} & \textbf{0.2644} \\ \hline
man-\textcolor{brightpink}{pink}-\textcolor{darkwhite}{white} & 0.2425 & 0.2400 & 0.2350 \\ \hline
\textbf{woman-black-\textcolor{metal}{metal}} & \textbf{0.2850} & \textbf{0.2675} & \textbf{0.2643} \\ \hline
\textbf{woman-\textcolor{brown}{brown}-\textcolor{green}{green}} & \textbf{0.2325} & \textbf{0.2184} & \textbf{0.2135} \\ \hline
\end{tabular}
}
\end{table}
%

\section{Prediction model}
The previous section demonstrates how several factors serve as indicators of popularity of the fashion boards on Pinterest. In this section, we shall leverage these factors to predict the future popularity of fashion boards. The popularity of a board is governed by the number of followers it has. To prevent any from of data leakage we separately re-crawl the new follower counts of all the fashion boards in our dataset in the month of April, 2017. This follower count statistics therefore is at a distance of 12 months from the training data. \footnote{Note that we do not use temporal statistics between March 2016 and April 2017 for enhanced robustness of the model; the idea is to make efficient predictions using minimal set of features that can be easily obtainable at any static time point.}

For the prediction task, we shall use the following features each of which is motivated by the analysis in the previous section.
\begin{compactitem}
\item Originality score;
\item Likes retention coefficient; 
\item Repins retention coefficient;
\item Likes production coefficient; 
\item Repins production coefficient;
\item Total number of pins;
\item Avg. no. of likes on pins; 
\item Avg. no. of repins of pins;
\item Avg. no. of comments on pins;
\item Inter-copying time; 
\item Duration of sharing;
\item Likes speed coefficient; 
\item Repins speed coefficient;
\item Gender counts (2 bins); Gender bi-term count;
\item Fashion term count (10 bins); Fashion bi-term count (10 bins); Fashion tri-term count (10 bins);
\item Color term count (10 bins); Color bi-term count (10 bins); %
\item Gender infused fashion bi-term count (10 bins); Gender infused fashion tri-term count (10 bins); Gender infused color count (5 bins). %
\end{compactitem}

\subsection{Predicting the popularity class of the boards} We have seen that the factors we have discussed earlier highly discriminate the unpopular class (LFB) from the two popular classes (HFB and ERB). The factors, however, can only moderately discriminate one of the popular class (HFB) from the other (ERB). We consider equal number of data points for each of the classes and then perform a 10-fold cross validation for generating results. In table~\ref{class}, we present the classification results for i) HFB vs LFB ii) ERB vs LFB and iii) HFB vs ERB. As evident from the table, we can discriminate both popular (HFB or ERB) from the unpopular class (LFB) very well with a very high accuracy (95.96\% for HFB vs LFB and 93.95\% for ERB vs LFB) and very high precision, recall and area under ROC curve. Note that we are only able to obtain a moderate accuracy (65.1\%) in classifying the two popular classes since the boards belonging to these two classes have very similar characteristics. 

For the classification task, we have used Support Vector Machines (SVM), Logistic Regression (LR) and Random Forest (RF) classifiers implemented in the Weka Toolkit~\cite{weka}. We choose the three
classifiers for their diversity since they are known to be able to
solve a vast range of different types of classification problems.
Each of these classifiers represent different schools of thoughts
and have their own set of strengths and advantages\footnote{https://bit.ly/2LkuSf0}. All the classifiers yield similar performance results with Random Forest classifier performing the best.
\begin{table}[h]
\small
\centering
\caption{Performance of various classifiers for classification of i) HFB vs LFB ii) ERB vs LFB ii) ERB vs HFB.}
\label{class}
 \begin{tabular}{ |p{1cm}|p{0.8cm}|p{1cm}|p{0.6cm}|p{0.6cm}|p{0.6cm}|p{0.6cm}|}
\hline
Categor-ies & Classif-iers & Accu-racy &Preci-sion & Recall & F-Score & ROC Area \\ \hline
\multirow{3}{1cm}{HFB vs LFB} & SVM & 92.51\% & 0.935  &  0.925  &  0.925 & 0.928\\\cline{2-7}
  &  LR & 94.91\% & 0.95  &  0.949  &   0.949  &  0.981\\ \cline{2-7}
  &  \cellcolor{pink}\textbf{RF} & \cellcolor{pink}\textbf{95.96\%} & \cellcolor{pink}\textbf{0.96}  & \cellcolor{pink}\textbf{0.96} & \cellcolor{pink}\textbf{0.96}  & \cellcolor{pink}\textbf{0.995 }\\ \hline
  \multirow{3}{1cm}{ERB vs LFB} & SVM & 92.06\% & 0.921&  0.921 & 0.921 & 0.92\\\cline{2-7}
  &  LR & 93.27\% & 0.934 & 0.933 & 0.933 & 0.977\\ \cline{2-7}
  &  \cellcolor{pink}\textbf{RF} & \cellcolor{pink}\textbf{93.95\%} & \cellcolor{pink}\textbf{0.94} & \cellcolor{pink}\textbf{0.94}  & \cellcolor{pink}\textbf{0.94} & \cellcolor{pink}\textbf{0.99} \\ \hline
    \multirow{3}{1cm}{ERB vs HFB} & SVM & 61.08\% & 0.62 & 0.611 & 0.603 & 0.611 \\\cline{2-7}
  &  \cellcolor{pink}\textbf{LR} & \cellcolor{pink}\textbf{65.1\%} & 0.655 & \cellcolor{pink}\textbf{0.651} &  \cellcolor{pink}\textbf{0.649}  & \cellcolor{pink}\textbf{0.684}\\ \cline{2-7}
  &  RF & 64.81\% & \cellcolor{pink}\textbf{0.672} & 0.648 & 0.636 & 0.625 \\ \hline
  \end{tabular}
\end{table}
\vspace{-2mm}
\subsection{Predicting the followership counts of the boards}
To study the robustness of our prediction model, we further try to predict the actual popularity, i.e., the logarithmic values of the followership counts of the boards. Toward this objective, we use Support Vector Regression (SVR) due to non-linearity of the problem. We use sequential minimal optimization (SMO) algorithm for training the SVR. We perform both separate training and testing as well as 10-fold cross validation method. We consider Pearson VII function-based universal kernel (PUK) due to its flexibility and adaptability through adjusting kernel parameter. We set the cost parameter (C) as 1. For evaluating how good the prediction is, we use Pearson correlation coefficient ($\rho$), normalized root mean square error (RMSE). We achieve high correlation coefficient ($0.8738$) and low normalized root mean square error ($0.1427$) which establishes the fact that the features obtained are robust and discriminating in nature (see table~\ref{regress}\footnote{We have also tried linear regression model which gives correlation coefficient of 0.7363 and 0.2 as normalized the RMSE value for 10-fold cross validation setting.}). Both cross-validation and separate training/testing produces very similar results. 
\begin{table}[h]
\centering
\vspace{-2mm}
\caption{Regression results.}
\vspace{-2mm}
\label{regress}
\resizebox{8cm}{!}{
\begin{tabular}{|p{4cm}|c|c|}
\hline
Method & $\rho$ & Normalized RMSE\\ \hline
10-fold cross-validation & 0.8659 & 0.146 \\ \hline
\cellcolor{pink}Separate training/testing (4:1 ratio) & \cellcolor{pink}\textbf{0.8738} & \cellcolor{pink}\textbf{0.1427}  \\ \hline
\end{tabular}}
\end{table}

\noindent\textbf{Discriminative features:} In order to determine the discriminative power of each feature, we use the $RELIEFF$ feature selection algorithm~\cite{relief} available in the Weka Toolkit. Table~\ref{tab:discri} shows the rank of the features in terms of their discriminating power for prediction. The rank order clearly indicates that for popularity prediction the sharing/borrowing features, the color terms and some of the fashion terms are important. Likes retention coefficient, repins retention coefficient are the top discriminative features followed by various color term based features. Therefore, color (sometimes in conjunction with fashion and gender term) seems to be one of the most important discriminator for popularity of fashion boards. 
\begin{table}[t]
\small
\centering
\vspace{-2mm}
  \caption{Top predictive features and their ranks.}
  \vspace{-2mm}
\label{tab:discri}
 \begin{tabular}{ |c|c| }
\hline
 Rank & Features\\ \hline
1 &  \cellcolor{pink}LRC  \\
2 &  \cellcolor{pink}RRC  \\
 3 & \cellcolor{pink}bag-striped (fashion)  \\
 4 & \textcolor{darkwhite}{white} (color)  \\
5 & black (color) \\
 6 & \textcolor{blue}blue (color)  \\
7 & female  \\
8 & male-female \\
9 & \textcolor{brown}{brown} (color) \\
10 & \textcolor{blue}{blue}-\textcolor{brightpink}{pink} (color) \\
11 & \textcolor{brightpink}{pink} (color) \\
12 & black-\textcolor{brightpink}{pink} (color) \\
13 & woman-\textcolor{brightpink}{pink}-\textcolor{darkwhite}{white} (gender-color) \\
14 & \textcolor{red}{red} (color) \\
15 & man-shirts-shorts (gender-fashion) \\\hline
\end{tabular}
\end{table}
\vspace{-2mm}

\section{Discussions and conclusions}
In this section we outline various insights and implications of the current work. We also discuss the generalizability of the current work and finally draw the conclusions.
\vspace{-2mm}
\subsection{Insights and implications}
\noindent\textbf{\em Insights}: The current study puts forward a lot of insights especially for new and upcoming fashion brands. 
\begin{compactitem}
\item Certain social sharing behavior of users can make boards popular. The most crucial among these are the retention coefficients. Popular boards are able to retain their proportion of `likes' and `repins' despite a lot of sharing and re-sharing of pins. Specially, engineered campaigns by the fashion houses can be made to ensure/promote such retentions.
\item More female faces or both male and female faces together may be promoted by the fashion houses since that, as we have seen, could lead to enhanced popularity.
\item Certain choices of colors (e.g., \textcolor{darkwhite}{white}, black, \textcolor{blue}{blue}, \textcolor{brightpink}{pink} etc.) and color combinations (e.g., \textcolor{blue}{blue}-\textcolor{brightpink}{pink}, black-\textcolor{brightpink}{pink} etc.) may be more advertised to enhance the chances of being more popular.
\item Certain fashion items like `striped bags' seem to be very common in popular brands and could be more promoted by the newbies.
\item For male fashion, `shirts' and `shorts' are the items that seem to propel popularity and can therefore be more vigorously advertised by the new outlets. Many articles\footnote{https://www.cosmopolitan.com/sex-love/a63297/things-hot-guys-wear/}, in fact, have noted that shorts like boxers and bathing suits that end above the knee enhance the sex appeal of men.
\item For female fashion, colors like \textcolor{brightpink}{pink} and \textcolor{darkwhite}{white} seem to be good indicators of popularity. In fact, \textcolor{brightpink}{pink} has been the most favorite color of garments for women for a very long time\footnote{https://www.racked.com/2015/3/20/8260341/pink-color-history}. These therefore can be items of more focused publicity by the upcoming fashion agencies.
\end{compactitem}

\noindent\textit{\textbf{Implications}}: Our findings make several contributions to existing research. We believe, this research opens new pathway to understand new factors like colors, faces, fashion terms which are influential for understanding popularity. Our work also echoes some of the previous findings on impact of color on diffusion. We also suggest color combinations that makes a board popular. 
For newbie fashion houses and fashion trend-setters, our findings shed light on how images can be constructed so that they become popular. Pins of certain colors, more female faces or male-female joint faces could be some of the prime suggestions. One could also launch campaigns to promote their boards in such a way that the originating boards are able to retain the `likes' and `repins' of their pins in the face of constant sharing of these pins.
In fact, Pinterest can make such `tips-n-tricks' application available in exchange of a small amount of subscription from every newbie. This could potentially be a premium/paid support and could be a business model for the company for possibility of enhanced revenues.

There are several mobile apps which provide users with photo-editing tools. One of the widely used techniques in photo-editing is applying filters to them. These filters can change saturation, brightness, and color distribution of the image. Our findings can be used to design new filters for photo editing. Filters
that increase saturation or enhance the warmness of the image will likely increase engagement with the photo.
\vspace{-3mm}
\subsection{Generalizability} Though the entire study has been performed on Pinterest, the findings can be generalized in other similar websites focused on images, for example, professional photography site like Flickr, or people-focused website like Instagram. Instagram is also a quite popular website for fashion trend. We believe these findings in the form of importance of color combinations and fashion terms influencing popularity can be generalized to Instagram, Flickr and Tumblr as well, though the popularity figures might vary which is mostly dependent on the website's underlying usage among communities, ranking algorithms etc.

\subsection{Conclusions}
In summary, we study various aspects of fashion boards on Pinterest. Our proposed measures -- retention coefficients, production coefficients, inter-copying time and duration of sharing portray the sharing dynamics evident in less followed, highly followed and expert ranked fashion boards.  

We observe that generally highly followed and expert ranked fashion boards are able to make an existing non-popular pin popular, whereas less popular fashion boards do not exhibit this characteristic. Further, if a pin has originated from highly followed or expert ranked fashion boards, it would achieve high popularity on this board than the subsequent boards on which it would be shared in future. We also observe that the pins on the highly followed and expert ranked fashion boards keep getting shared for a long time, whereas this happens for a short time for the pins on less followed fashion boards.

Gender, fashion and color terms embedded in images also yield interesting and conclusive results. We observe that both highly followed and expert ranked fashion boards exhibit similar trend in the usage of fashion bi- and tri-terms. We also observe that a large number of pins having female faces are present in highly followed and expert ranked fashion boards; the number of female faces is 20\% lower for the less followed boards. Similar trend is observed for pins having both male and female faces. We also study occurrences of gender-based fashion and color terms. We identify combinations which give good discriminatory results across the three board categories. We try to leverage various sharing/borrowing characteristics, image-based content characteristics of fashion boards to predict their future popularity (logarithm of follower count). We achieve a high correlation coefficient of $0.874$ and low RMSE.

\noindent\textit{Limitations}: We acknowledge that there is some limitation of the current study. We specifically note the fact that some of the features and outcomes might be influenced by the particulars of the Pinterest ranking algorithms (e.g., what gets featured on the homepage, how personalization affects the probability a pin will be surfaced, etc.). There is no way we can control the internal algorithm promoting pins and boards. However, we believe, the factors we come up with are indeed influential as they strongly correlate with popularity studied on a large-scale data. 
\bibliographystyle{aaai}
\bibliography{Bibliography}
\end{document}